\documentclass[floatfix,preprint,aip,jap]{revtex4-2} 
\usepackage{lipsum}
\usepackage{graphicx}
\usepackage{amsmath}
\usepackage{amssymb}
\usepackage{epsfig}
\usepackage{subeqnarray}
\usepackage{color}
\usepackage{bbold}
\usepackage{dsfont}
\usepackage{tikz}
\usepackage{cancel}
\usetikzlibrary{positioning,calc}
\usepackage[version=3]{mhchem}
\usepackage{color}
\usepackage[normalem]{ulem}

\newcommand{\fig}[1]{Figure~\ref{#1}}

\newcommand{\degree}{\ensuremath{^\circ}}

\newcommand{\ie}{\textit{i.e.},~}

\newcommand{\eg}{\textit{e.g.,~}}
\newcommand{\etal}{\textit{et al.~}}

\newcommand{\gaox}{\ce{\beta-Ga_2O_3}}
\graphicspath{{./}}
\begin{document}

\title{Anisotropic Anharmonicity Dictates the Thermal Conductivity of Gallium Oxide (\ce{\beta-Ga_2O_3})}

\author{Abdulaziz Alkandari}
\affiliation{School of Mechanical Engineering and Birck Nanotechnology Center, Purdue University, West Lafayette, 47907, IN, USA}
\affiliation{Mechanical Engineering Department, College of Engineering and Petroleum, Kuwait University, PO Box 5969, Safat 13060, Kuwait}
\author{Zherui Han}
\affiliation{School of Mechanical Engineering and Birck Nanotechnology Center, Purdue University, West Lafayette, 47907, IN, USA}
\author{Ziqi Guo}
\affiliation{School of Mechanical Engineering and Birck Nanotechnology Center, Purdue University, West Lafayette, 47907, IN, USA}
\author{Thomas E. Beechem}
\email{tbeechem@purdue.edu}
\affiliation{School of Mechanical Engineering and Birck Nanotechnology Center, Purdue University, West Lafayette, 47907, IN, USA}
\author{Xiulin Ruan}
\email{ruan@purdue.edu}
\affiliation{School of Mechanical Engineering and Birck Nanotechnology Center, Purdue University, West Lafayette, 47907, IN, USA}
\date{\today}

%-------------------------------Abstract----------------------------------------%

\begin{abstract}
\hfill

\ce{\beta-Ga_2O_3} is a promising material candidate for next-generation high power devices even as its low thermal conductivity (\ce{\kappa}) limits utilization due to an inability to sufficiently dissipate heat. Despite the importance of this inherent thermal challenge, a significant discrepancy persists between experimental results and computational models regarding \ce{\beta-Ga_2O_3}'s anisotropic thermal conductivity. Specifically, computational results are within experimental error bounds for \ce{\kappa_100} and \ce{\kappa_001} while underpredicting \ce{\kappa_010}, suggesting that the bare phonon models used in literature are missing essential physics related to the anisotropic thermal transport. In response, we compute the anisotropic \ce{\kappa} using first-principles and the Pierels-Boltzmann transport equation (PBTE) under different approximations. For the simplest model, we consider the heat carriers to be harmonic phonons with scattering rates obtained perturbatively. These results are then compared to those obtained by including phonon renormalization and four-phonon scattering. Our results show that accounting for phonon renormalization resolves the discrepancy between experiment and theory. This is because phonon renormalization leads to an anisotropic \ce{\kappa} enhancement caused by directionally-dependent changes in the phonon group velocities accompanied by a general increase in phonon lifetime. Owing to the crucial role of these anharmonic interactions in accurately describing anisotropic thermal transport, we also explore the anharmonicity of individual atoms and show that the octahedrally-coordinated gallium atom is the most anharmonic and thus most likely responsible for the failure of the harmonic phonon model to describe thermal transport in this material. Finally, we demonstrate that atomic anharmonicities could be used as a useful metric to guide the tailoring of vibrational properties.
\end{abstract}
\maketitle
%------------------------------------------------------------------------------%

%-------------------------------Introduction-------------------------------------%
\section{Introduction}

Ultrawide-bandgap ($>$ 3.4 eV) semiconductors have reached a technological maturity in which fundamental potential has motivated significant effort pursuing commercial adoption.\cite{tsao_2018} Since the breakdown voltage increases nonlinearly with the bandgap, ultrawide-bandgap materials are promising candidates for robust power device performance at high voltages and temperatures.\cite{wong_2021} To capitalize on this potential, fundamental understanding of the material physics in those regimes must be known.  There remains, however, knowledge gaps regarding even intrinsic material properties like thermal conductivity in these materials.\cite{beechem_2016,song_2021}  With this motivation, we examine here the magnitude and underlying mechanisms belying the highly anisotropic nature of thermal conductivity in \ce{\beta-Ga_2O_3}. 

\ce{\beta-Ga_2O_3} is an ultrawide-bandgap (4.8 eV)\cite{sasaki_2013} material that is commercially-available in high-quality wafers produced using inexpensive melt growth processes.\cite{higashiwaki_2014} This allows \ce{\beta-Ga_2O_3} power devices to be more economically viable than other wide-bandgap materials such as SiC and GaN.\cite{reese_2019} However, the thermal conductivity (\ce{\kappa}) of \ce{\beta-Ga_2O_3} is very low (11 - 27 W/m $\cdot$ K)\cite{guo_2015} when compared to these materials (\ce{\kappa_S_i_C} $\approx$ 500~W/m $\cdot$ K~\cite{cheng_2022}, \ce{\kappa_G_a_N} $\approx$ 230 W/m $\cdot$ K\cite{mion_2006}). The low \ce{\kappa} necessitates more aggressive thermal management schemes, such as double-side cooling.\cite{kim_2023} Therefore, thermal management is one of the significant barriers towards commercialization of \ce{\beta-Ga_2O_3} electronics. A robust understanding of its intrinsic thermal transport properties is, therefore, warranted.

\begin{figure}[htbp]
\centering
\includegraphics[scale = 0.3]{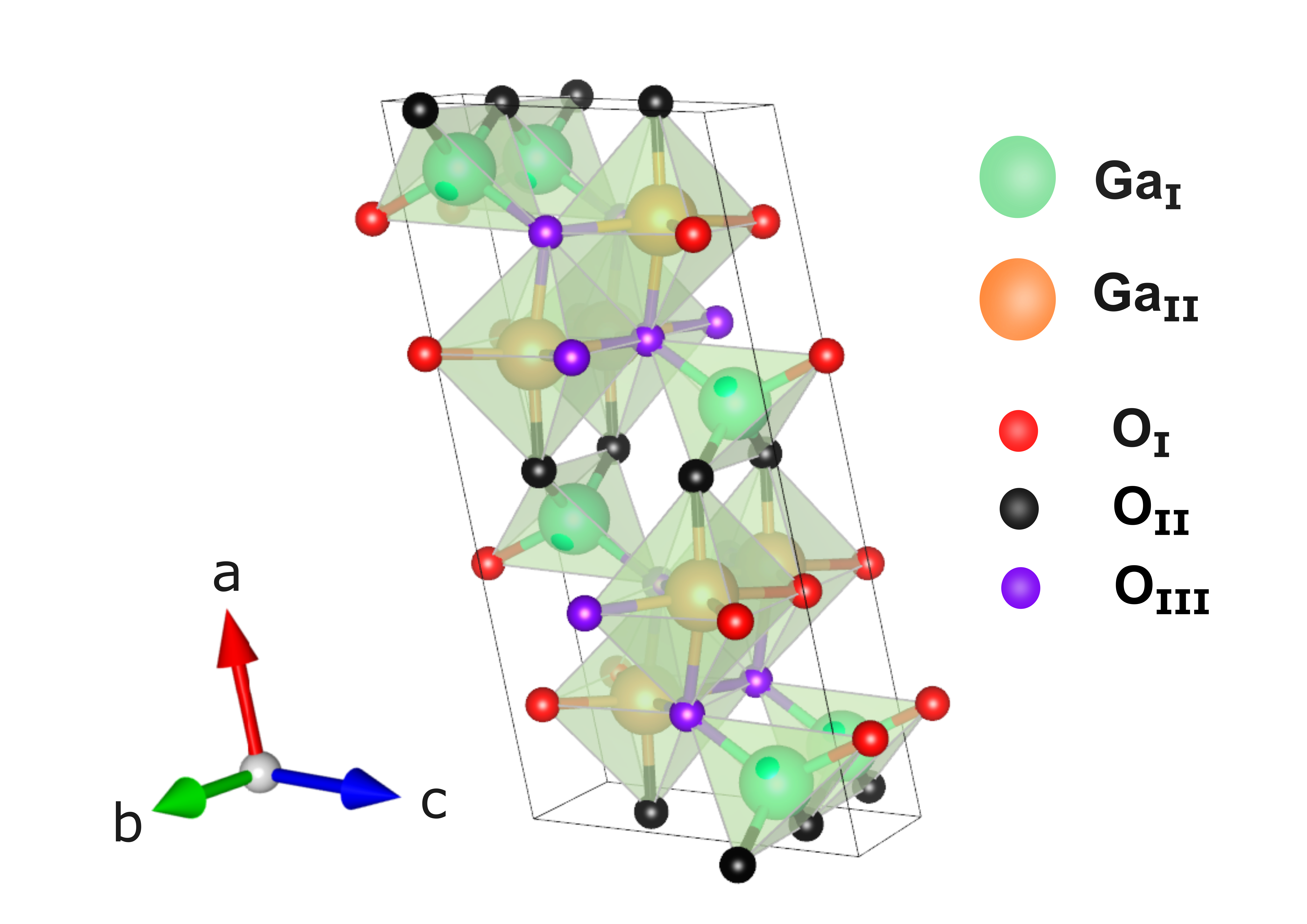}
\caption{The conventional unit cell of monoclinic \ce{\beta-Ga_2O_3}. The unit cell contains 20 atoms with 5 crystallography nonequivalent atoms. One gallium atom is tetrahedrally-coordinated (\ce{Ga_I}) while the other is octahedrally-coordinated (\ce{Ga_I_I}). The three remaining oxygen atoms sit between the polyhedra: (\ce{O_I}) is at the corner of two octahdra and one tetrahedron, (\ce{O_I_I}) is at the corner of two tetrahdra and one octahedron, and (\ce{O_I_I_I}) is at the corner of three octahedra and one tetrahdron.\cite{ahman_1996}   }
\label{Fig_1}
\end{figure}

\ce{\beta-Ga_2O_3} has a monoclinic crystal structure, as shown in \fig{Fig_1}, leading to four independent components of the thermal conductivity tensor.\cite{newnham_2005} The crystal structure is also highly anisotropic (a $=$ 12.214 \r{A}, b $=$ 3.0371 \r{A}, c $=$ 5.7981 \r{A}, \ce{\beta} $=$ 103.83\degree )\cite{ahman_1996}, causing significant anisotropy in the measured thermal conductivity (\ce{\kappa_100} = 9.5-18, \ce{\kappa_010} = 22.5-29.0, \ce{\kappa_001} = 12.7-21.0 W/m $\cdot$ K).\cite{guo_2015,handwerg_2016,jiang_2018,slomski_2017,klimm_2023} The highly anisotropic thermal conductivity of \gaox ~ differentiates it from other ultrawide bandgap materials like GaN that effectively transports heat isotropically despite possessing a thermal conductivity tensor that is formally anisotropic.\cite{lindsay_2012} 

The thermal conductivity of \ce{\beta-Ga_2O_3} has been extensively studied both experimentally\cite{guo_2015,jiang_2018,handwerg_2016,slomski_2017,klimm_2023} and computationally.\cite{santia_2015,yan_2018} However, discrepancies exist between the computational and experimental thermal conductivities. Specifically, the computational results are within experimental error bounds for \ce{\kappa_100} and \ce{\kappa_001} while under-predicting \ce{\kappa_010} ($\approx$ 4 - 32 \%). This peculiar mismatch suggests that computational models may be missing important physics relevant to the anisotropic thermal transport in \ce{\beta-Ga_2O_3}.

The combination of Density Functional Theory (DFT) and Peirels-Boltzmann Transport Equation (PBTE) has led to a framework capable of predicting thermal transport in solids with high accuracy.\cite{lindsay_2019} However, differences between the experimentally measured \ce{\kappa} and those obtained from the standard computational workflow (described in detail in Ref. \cite{mcgaughey_2019}) do exist when the assumptions inherent to the predictions conflict with the reality of the system under study. More simply, discrepancies between experiment and theory happen when assumptions break down.  It is, therefore, necessary to consider the assumptions underlying predictions of the anisotropic thermal conductivity in \gaox.

For \gaox, the standard PBTE method, which presumes that the phonons are well-defined quasiparticles, is valid. This assumption is justified if the phonon mean free paths are reasonably larger than the interatomic spacing, or if the phonon lifetimes are greater than the inverse frequency of the phonon mode.\cite{simoncelli_2022}
For \ce{\beta-Ga_2O_3}, experimental measurements of the phonon lifetimes\cite{schubert_2016} are consistent with this picture. In addition, \gaox ~exhibits an appreciably larger thermal conductivity relative to materials where the formalism is known to break down.\cite{knoop_2023} Taken together, the phonon quasiparticle picture is reasonable to use with \gaox ~and hence the PBTE is valid. Therefore, we turn our attention onto the approximations introduced by DFT and the difficulties that come about when using it to accurately predict thermal conductivity along all directions in \gaox. 

Defects are an obvious candidate. Standard DFT approaches employ periodic boundary conditions where the crystal structure is assumed to be defect-free. This is in contrast to the imperfections that exist in almost every ``real-world" material. Defects are not, however, the chief cause here. For example, although neglecting phonon-defect scattering may explain why \ce{\kappa} is overpredicted in two crystal directions (\eg \ce{\kappa_100} and \ce{\kappa_001}), thermal conductivity is actually under predicted along \ce{\kappa_010} direction. It seems very unlikely that defects would mitigate transport along two directions while boosting it along the third. Therefore, this explanation is discarded.  

Instead attention is turned to the role of anharmonicity on anistropic thermal transport. DFT is an inherently ground-state approach.\cite{martin_2004} This limits any perturbative treatment of the lattice potential energy expansion terms to be temperature-independent. Therefore, obtaining the phonon energies from the Hessian matrix of the expansion, known as the harmonic approximation, leads to phonon energies that are also temperature-independent.\cite{baroni_2001} The finite-temperature effects on the potential energy expansion terms\textemdash and hence on the phonon energies and even the atomic displacement themselves (\ie polarizations)\textemdash are known as phonon renormalization effects. The extent to which these effects impact \ce{\kappa} in \ce{\beta-Ga_2O_3} in general, and along different crystal in particular, is unknown. Therefore, we focus on anharmonicity and how it manifests specifically in the thermal transport of \gaox.

Phonon renormalization effects are negligible for \ce{\kappa} at room temperature in strongly-bonded solids such as silicon, diamond, and MgO.\cite{ravichandran_2018,han_2023c} However, they are crucial for accurate prediction of \ce{\kappa} in other solids such as NaCl\cite{ravichandran_2018}, PbTe\cite{romero_2015}, and a wide range of oxides including \ce{TiO_2}\cite{fu_2022}, \ce{SrTiO_3}\cite{guo_2023}, and \ce{CeO_2}.\cite{han_2023c} Accounting for phonon renormalization in the calculation of \ce{\kappa} in \ce{TiO_2}, for example, leads to a 50\% increase at room temperature and significantly better agreement with experimental data.\cite{fu_2022} Therefore, phonon renormalization could play a significant role in determining the thermal conductivity of \gaox ~and, if ignored, lead to discrepancies between experiment and prediction. 

Similarly, four-phonon (4-phonon) scattering may also be significant, requiring its assessment as well.  Higher-order phonon scattering can significantly reduce \ce{\kappa}, \cite{feng_2016,feng_2017,yang_2019,chen_2023a,guo_2024a} especially in materials whose optical phonons significantly contribute to the thermal transport.\cite{yang_2019} Previous work suggested that optical phonon modes contribute more than 50\% to \ce{\kappa_010} in \gaox.\cite{santia_2015} Recognizing this, we examine here the the net interplay between phonon renormalization and four-phonon scattering and its impact on the anisotropic response of thermal conductivity in \gaox. 

Specifically, we combine a self-consistent phonon framework with a solution of the PBTE to include phonon renormalization, thermal expansion, and four-phonon scattering to calculate \ce{\kappa} of \ce{\beta-Ga_2O_3} in the temperature range between 200 and 600 K. The upper limit of this range corresponds to the operational temperature in electric vehicles, where the temperature near the engine can reach up to 200\degree C\cite{huque_2009}. We observe that including phonon renormalization resolves the discrepancy between experimental and computational \ce{\kappa} in \ce{\beta-Ga_2O_3}. The \ce{\kappa} enhancement due to renormalization is anisotropic and highest for \ce{\kappa_010}; the same tensor element that was significantly underpredicted by previous computational studies. Furthermore, we show that four-phonon scattering is weak relative to the dominant three-phonon scattering in this temperature range. By quantifying the anharmonicity of the individual atoms, we discover that the motion of the octahedrally-coordinated gallium atom (\ce{Ga_I_I}) is the most sensitive to the temperature-dependent potential energy surface (\ie is the most anharmonic). Practically, this suggests that enhancing \ce{\kappa} of \ce{\beta-Ga_2O_3} could be achieved by alloying with an element that takes the \ce{Ga_I_I} position in the crystal structure.

%------------------------------------------------------------------------------------------------%
\section{Methodology}

In this section, we elaborate on our workflow to calculate \ce{\kappa} of \ce{\beta-Ga_2O_3} using a first-principles approach. From a solution of the PBTE for the phonon equilibrium distribution function, $f_{{\lambda}}$,  the thermal conductivity tensor\cite{li_2014} is expressed as:

\begin{equation}
\label{Eq_1}
\kappa^{\alpha\beta} = \frac{1}{k_B T^2 N V} \sum_{\lambda} (\hbar\omega_\lambda)^2 f_{{\lambda}}^{0} ( f_{{\lambda}}^{0} + 1) \mathbf{v}_{\lambda}^{\alpha} \mathbf{v}_{\lambda}^{\beta} \tau_{\lambda}
\end{equation}

\noindent where $\lambda$ is a phonon mode index including polarization and wave vector,
$\alpha$ and $\beta$ are Cartesian directions, $k_B$ is the Boltzmann constant, $N$ is the number of unit cells, $V$ is the volume of the unit cell, $\hbar$ is the reduced Planck constant, $\omega_\lambda$ is the mode frequency, $\mathbf{v}_{\lambda}$ is the mode group velocity, and $\tau_{\lambda}$ is the mode lifetime. Therefore the evaluation of  \ce{\kappa} requires knowledge of the energy carrier (phonons) velocities and lifetimes. 

These velocities and lifetimes are obtained from DFT using either the ground-state (standard approach) or from a finite-temperature (effective) potential energy surface (PES).
To understand this in more detail, we begin by writing the Taylor expansion of the lattice potential energy ($U$) about the equilibrium position of the atoms:

\begin{equation}
\label{Eq_2}
U = U_0 + \frac{1}{2!} \sum_{ij} \frac{\partial^2 U}{\partial \mathbf u_i \partial \mathbf u_j} \mathbf u_i \mathbf u_j + \frac{1}{3!} \sum_{ijk} \frac{\partial^3 U}{\partial \mathbf u_i \partial \mathbf u_j \partial \mathbf u_k} \mathbf u_i \mathbf u_j \mathbf u_k + \frac{1}{4!} \sum_{ijkl} \frac{\partial^4 U}{\partial \mathbf u_i \partial \mathbf u_j \partial \mathbf u_k \partial \mathbf u_l} \mathbf u_i \mathbf u_j \mathbf u_k \mathbf u_l + \dots
\end{equation}

\noindent where $U_0$ is the energy at the equilibrium position, $\mathbf u_i$ represents the displacement of atom $i$ from its equilibrium position, and the derivatives represent increasing orders of interatomic force constants (IFCs). The second-order IFC (Hessian) is used to calculate the phonon dispersion and therefore velocities. This is done by diagonalization of the dynamical matrix assembled from the mass-normalized Hessian. The higher-order IFCs are needed to calculate the three and four-phonon scattering rates. %Following the mathematical formulation, we now attempt to build a mental picture of the problem. 

Terminating the expansion at the Hessian is known as the harmonic approximation. In this approximation, the lattice energy can be written as a sum of non-interacting normal modes. This assumption is necessary to obtain the phonon energies as including higher-order IFC into the lattice potential energy no longer leads to a diagonalizable dynamical matrix. This makes accounting for anharmonicity in the phonon energies (\ie phonon renormalization) problematic. 
The Temperature-Dependent Effective Potential (TDEP) accomplishes this task by introducing model IFCs that represent an effective potential energy surface (similar to Equation 2). This effectively maps the effect of phonon scattering onto the model Hessian. This model Hessian can be diagonalized to obtain the phonon energies renormalized by phonon scattering, leading to temperature-dependent phonon energies. To do this, TDEP constructs the model Hessian, and the higher order IFCs, by minimizing the difference between the model forces and force-displacement datasets.

In this study, we utilize the stochastic Temperature Dependent Effective Potential (s-TDEP)\cite{hellman_2013,hellman_2013b,hellman_2011,knoop_2024} to extract the renormalized Hessian and higher order IFCs. The s-TDEP method is based on stochastic generation of atomic thermal displacements followed by fitting force-displacement datasets to a model Hamiltonian.\cite{shulumba_2017} Therefore, s-TDEP allows extracting the temperature-dependent PES, up to fourth-order terms and including zero-point quantum motion, at a reasonable computational cost compared to \textit{ab initio} molecular dynamics (AIMD). Up to this point, we showed how to obtain the phonon energies using either the ground-state Hessian, obtained directly from DFT, or from the temperature-dependent Hessian constructed by TDEP. The final task is to lay the theory necessary to calculate the total phonon scattering rates.  

Using the third and fourth order IFCs, the three (3ph) and four (4ph) phonon scattering rates can be evaluated using Fermi's golden rule.\cite{li_2014}. In this study, isotopic scattering is excluded to focus exclusively on the intrinsic phonon scattering mechanisms arising from the material's anharmonicity. To obtain the total scattering rate, Mattheisen's rule is employed to sum all the contributions from the individual scattering channels\cite{ziman_2001,li_2014,han_2022}:

\begin{equation}
\label{Eq_4}
 \frac{1}{\tau_\lambda}=\frac{1}{N_q}\left(\displaystyle\sum^{(+)}_{\lambda'\lambda''}\Gamma^{(+)}_{\lambda\lambda'\lambda''}+\displaystyle\sum^{(-)}_{\lambda'\lambda''}\frac{1}{2}\Gamma^{(-)}_{\lambda\lambda'\lambda''}\right)
 +\frac{1}{N}\left(\displaystyle\sum^{(++)}_{\lambda'\lambda''\lambda'''}\frac{1}{2}\Gamma^{(++)}_{\lambda\lambda'\lambda''\lambda'''}+\displaystyle\sum^{(+-)}_{\lambda'\lambda''\lambda'''}\frac{1}{2}\Gamma^{(+-)}_{\lambda\lambda'\lambda''\lambda'''}+\displaystyle\sum^{(--)}_{\lambda'\lambda''\lambda'''}\frac{1}{6}\Gamma^{(--)}_{\lambda\lambda'\lambda''\lambda'''}\right)
\end{equation}

\noindent where $N_q$ denotes the total number of grid points used in the PBTE solution. Superscripts $( \pm )$ and $( \pm\pm )$ differentiate between various 3ph and 4ph scattering mechanisms. The 3ph processes include absorption ($\lambda+\lambda' \rightarrow \lambda''$), where two phonons combine, and emission ($\lambda \rightarrow \lambda' + \lambda''$), where one phonon decays into two. For the higher-order 4ph interactions, there are recombination processes ($\lambda+\lambda'+\lambda'' \rightarrow \lambda'''$), involving the merging of three phonons, redistribution ($\lambda+\lambda' \rightarrow \lambda'' + \lambda'''$), where two phonons are transformed into another pair, and splitting ($\lambda \rightarrow \lambda' + \lambda'' + \lambda'''$), where one phonon splits into three.

Since this investigation focuses on the effect of anharmonicity on \ce{\kappa} in \ce{\beta-Ga_2O_3}, it is useful to consider a quantitative measure of anharmonicity. Such a metric allows relating macroscopic properties\textemdash here we focus on thermal transport\textemdash to the individual nuclei dynamics. %Enabling anharmonicity tuning of materials to either increase or decrease \ce{\kappa} has important practical usage in microelectronics. 
This is further motivated by the complex nature of \ce{\beta-Ga_2O_3} having 5 nonequivalent atoms in the primitive cell (\fig{Fig_1}). Typically, anharmonicity is quantified by perturbatively mapping out the ground-state PES of all the unique atom pairs (bonds) in the crystal structure.\cite{yue_2017,wu_2021} However, this method is tedious as the dimensionality of the PES scales with the number of atoms in the unit cell. Furthermore, if we consider that the ground-state PES might not accurately describe the dynamics of nuclei at finite temperatures, then there is no guarantee that anharmonicity quantification using the ground-state PES yields valid results. More recently, Knoop \etal \cite{knoop_2020} derived a measure of anharmonicity ($\sigma_{\text{A}}$) based on the standard deviation of the average difference between the ground-state harmonic forces and AIMD forces:

\begin{equation}
\label{Eq_5}
 {\sigma_{\text{A}}}(T) = \sqrt{\frac{\sum_{I \alpha}\langle (F_{I \alpha} - F_{I \alpha}^{(2)})^2 \rangle}{\sum_{I \alpha}\langle (F_{I \alpha})^2 \rangle}}
\end{equation}
where $I$ represents the atom index and $\alpha$ represents a Cartesian direction. In this study, we employ a slightly different modification of their metric. Instead of using the ground-state harmonic forces, we use the temperature-dependent harmonic forces obtained from s-TDEP. This allows us to evaluate the anharmonicity of the individual atoms in the crystal structure at no additional computational cost. Anharmonicity is typically regarded as a material property.\cite{knoop_2020} Here, we examine how it manifests in very anisotropic materials at the level of individual atoms. 

We use VASP 5.4.1 \cite{kresse_1993,kresse_1996,kresse_1996c} for all DFT calculations,FourPhonon\cite{han_2022} for the PBTE solution, and TDEP\cite{hellman_2013,hellman_2013b,hellman_2011,knoop_2024} to extract the temperature-dependent IFC. The detailed numerical parameters are discussed in the Supplementary Materials.\cite{_bo} At the time of preparing the manuscript, we became aware of a study on \ce{\kappa} in \ce{\beta-Ga_2O_3} including the effect of phonon renormalization using TDEP.\cite{chen_2023} However, they showed that including phonon renormalization does not fully resolve underprediction discrepancy for \ce{\kappa_010}. Our study shows otherwise. We attribute this to their much coarser DFT numerical settings. For example, we use a cut-off energy of 640 eV with a (4$\times$4$\times$4) mesh for the supercell, while they use a 400 eV cut-off energy and a (1x1x1) mesh.

%------------------------------------------------------------------------------------------------%
\section{Results and Discussion}

\begin{figure}[htbp]
\centering
\includegraphics[scale = 0.65]{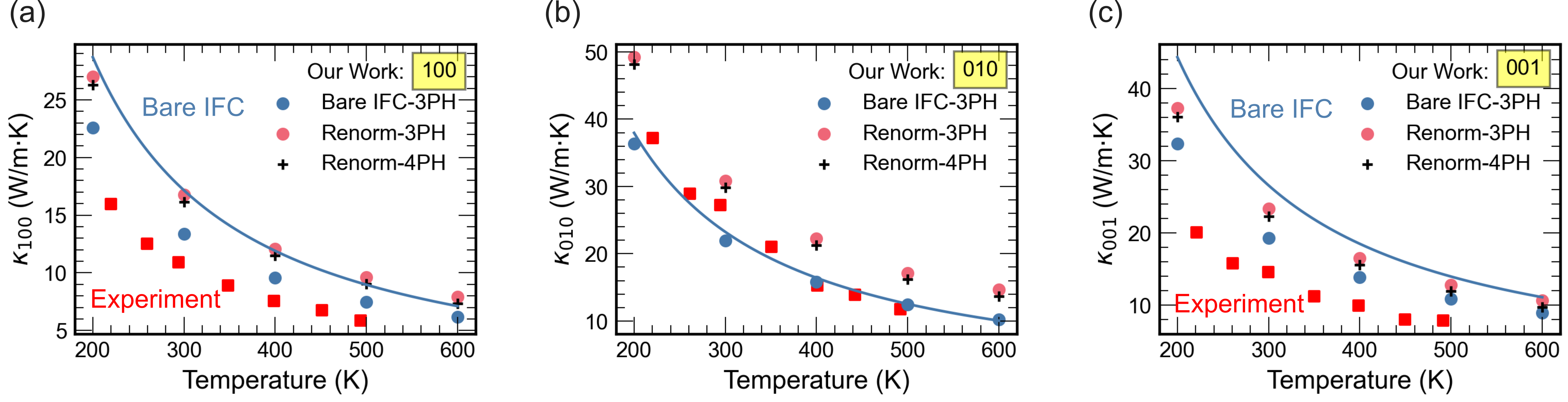}
\caption{Plots of the thermal conductivity \ce{\kappa} as a function of temperature for three directions: (a) \ce{\kappa_100}, (b) \ce{\kappa_010}, and (c) \ce{\kappa_001}. We include experimental measurements (red squares \cite{guo_2015}) and bare IFC calculation fits (blue line \cite{santia_2015}) for comparison. For our results, we include \ce{\kappa} calculated in different approximations: bare IFC with 3-phonon scattering (blue symbols), renormalized phonons (Renorm) with 3-phonon scattering (red symbols), and renormalized phonons with 3-phonon and 4-phonon scattering (black symbols).}
\label{Fig_2}
\end{figure}

In \fig{Fig_2}, we plot the temperature-dependence of the anisotropic thermal conductivity of \ce{\beta-Ga_2O_3} for each of the independent crystal directions. Discrepancies between experiment and theory are evident and can be mitigated with the inclusion of the salient phonon renormalization effects. Predictions of thermal conductivity using the bare IFCs from literature (blue curve\cite{santia_2015}) appreciably overpredict the experimental results (red squares\cite{guo_2015}) for \ce{\kappa_100} and \ce{\kappa_001}. However, for \ce{\kappa_010}, the bare IFCs prediction is nearly equivalent to experiment over a range of temperatures. It is unclear why defects "should" affect transport in one direction more than another. Furthermore, since the computational approach presumes no extrinsic defects while actual materials are beset by them, thermal conductivity predictions should be larger than their measured counterparts to some degree. For these reasons, we assert that the directional differences are due to some other factor beyond that of defects. 

\fig{Fig_2} also includes a summary of our calculated results including 3-phonon scattering with the bare IFCs, with phonon renormalization, and including 4-phonon scattering with phonon renormalization (blue,red, and black symbols respectively). For \ce{\kappa_100} and \ce{\kappa_001}, our bare IFC phonon results are slightly lower than the computational results from literature.
Once the effect of phonon normalization is introduced, \ce{\kappa_100} and \ce{\kappa_001} increase and are in very good agreement to the bare IFCs \ce{\kappa} from literature. 
We attribute differences between our bare IFC results and those from literature mainly to different supercell sizes (160 atoms in our work vs. 80) and exchange-correlation functional (PBEsol in our work vs. GGA).  

For \ce{\kappa_010}, our bare phonon calculations still anomalously overlap the experimental and computational literature values. However, once we introduce phonon renormalization into the picture, \ce{\kappa_010} increases and becomes in line with the observations of thermal conductivity in the other directions (\ie slightly higher than that of the experiment). This resolves the discrepancy between experiment and theory for \ce{\kappa_010}. Including four-phonon scattering has a small ($<$8 \%) effect on \ce{\kappa} in all directions and across the temperatures considered. This is because the phase space for 3-phonon processes is huge ($\approx \; 10^9$ processes). This leads to phonon lifetimes being dominated by 3-phonon scattering processes rather than the weaker 4-phonon processes, which typically become dominant if the 3-phonon phase space is restricted. Therefore, the following discussion probing the anisotropic nature of anharmonicity is performed including 3-phonon scattering only.

Up to this point, we have shown that including phonon renormalization in the computational model resolves the anisotropic discrepancy in \ce{\kappa} between theory and experiment. This is a case where anharmonicity manifests itself anisotropically into the material properties. Therefore, anharmonicity is not a binary metric for materials with low-symmetry. The importance of this for practical applications cannot be understated. The ability to tune a material's anharmonicity in a given direction may be crucial in applications where material anisotropy is relevant. For example, \ce{\kappa} anisotropy could be leveraged to control heat flow path in electronic systems\cite{dede_2010}. 

\begin{figure}[htbp]
\centering
\includegraphics[scale = 0.65]{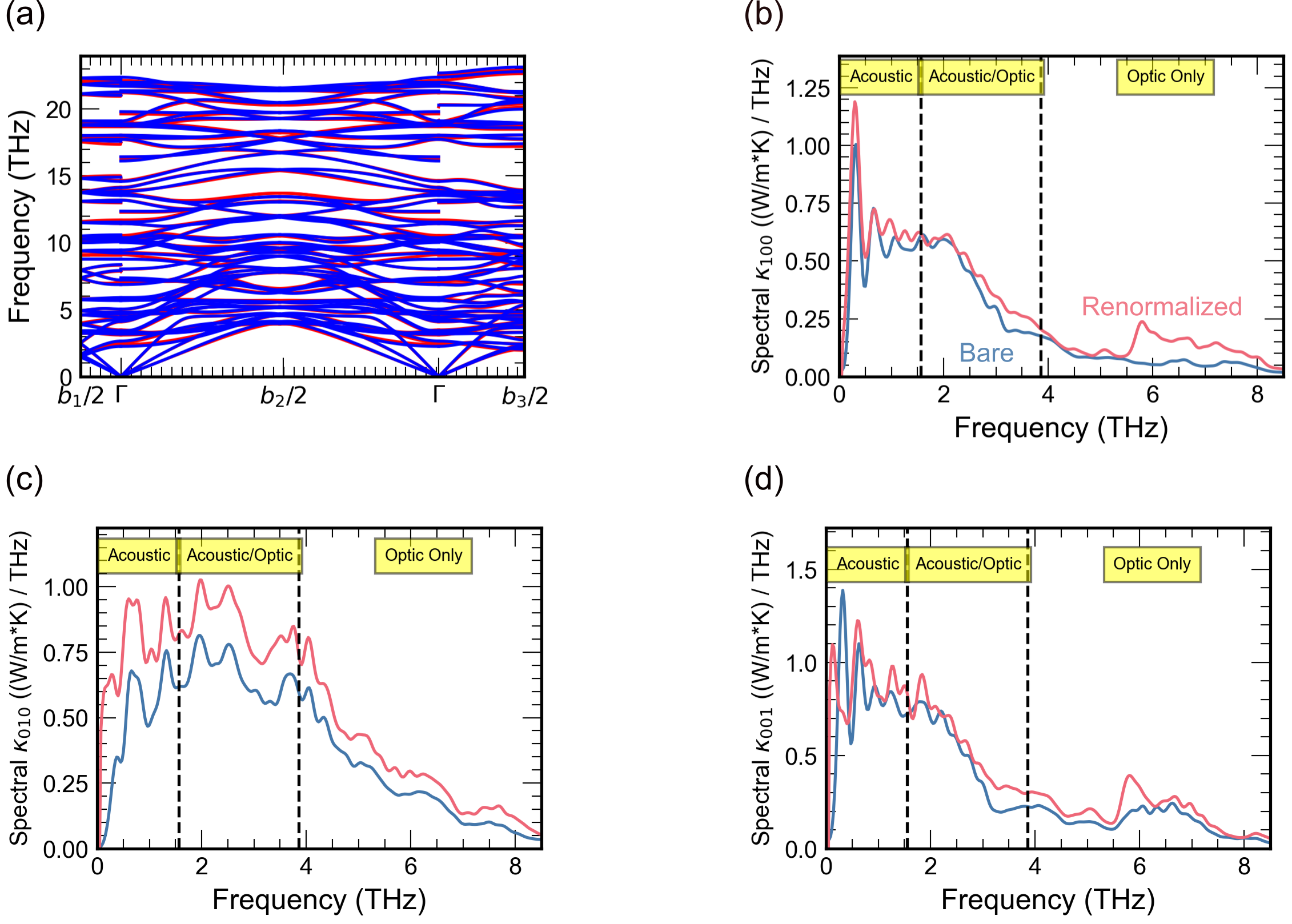}
\caption{Phonon dispersion (a) plotted along the three reciprocal lattice vectors for the bare phonon model (blue) and the model including phonon renormalization at 300 K (red). Spectral thermal conductivity for (b) \ce{\kappa_100}, (c) \ce{\kappa_010},and (d) \ce{\kappa_010} obtained at 300 K using both bare (blue) and renormalized phonons (red). The dashed vertical lines represent the frequency of the lowest optical mode and the highest acoustic mode respectively.}
\label{Fig_3}
\end{figure}

Now we turn our attention to understanding the microscopic origins by which phonon renormalization enhances \ce{\kappa} along certain directions more than others. As shown in \fig{Fig_3}(a), accounting for phonon renormalization results in a slight shift of the phonon mode energies. To further isolate the phonon modes responsible for the \ce{\kappa} enhancement, we plot the anisotropic spectral thermal conductivity at 300 K in \fig{Fig_3}(b-d) for both the bare IFC and renormalized phonons. We observe that across all directions, more than 90\% of the \ce{\kappa} enhancement is due to phonon modes under 8 THz. The spectral functions also show the anisotropic nature of the \ce{\kappa} enhancement due to phonon renormalization: 20\% for \ce{\kappa_100}, 29\% for \ce{\kappa_010}, and 17\% for \ce{\kappa_010}. Furthermore, decomposing \ce{\kappa} into acoustic and optical phonon contributions—as shown in \fig{Fig_3}(b–d)—reveals that both types of modes are carrying more heat with the inclusion of phonon renormalization. While inclusion of the effect modifies the relative contributions from acoustic and optical modes, the contribution from optical phonon modes to \ce{\kappa} remains significant. For example, optical modes are responsible for 48\% of the heat transport in \ce{\kappa_010}. This high optical phonon contribution is common in complex materials with significant \ce{\kappa} anisotropy.\cite{mukhopadhyay_2016,li_2021a,li_2021,pandey_2022}

Since we are dealing with bulk solids, phonon modes carry more heat either because of an increase in the group velocity or a decrease in the phonon scattering rate. Therefore, we investigate the changes in group velocity and lifetimes of the relevant phonon modes in an attempt to answer two questions: why does \ce{\kappa} increase when including phonon renormalization in the computational model and why is this increase anisotropic?

\begin{figure}[htbp]
\centering
\includegraphics[scale = 0.65]{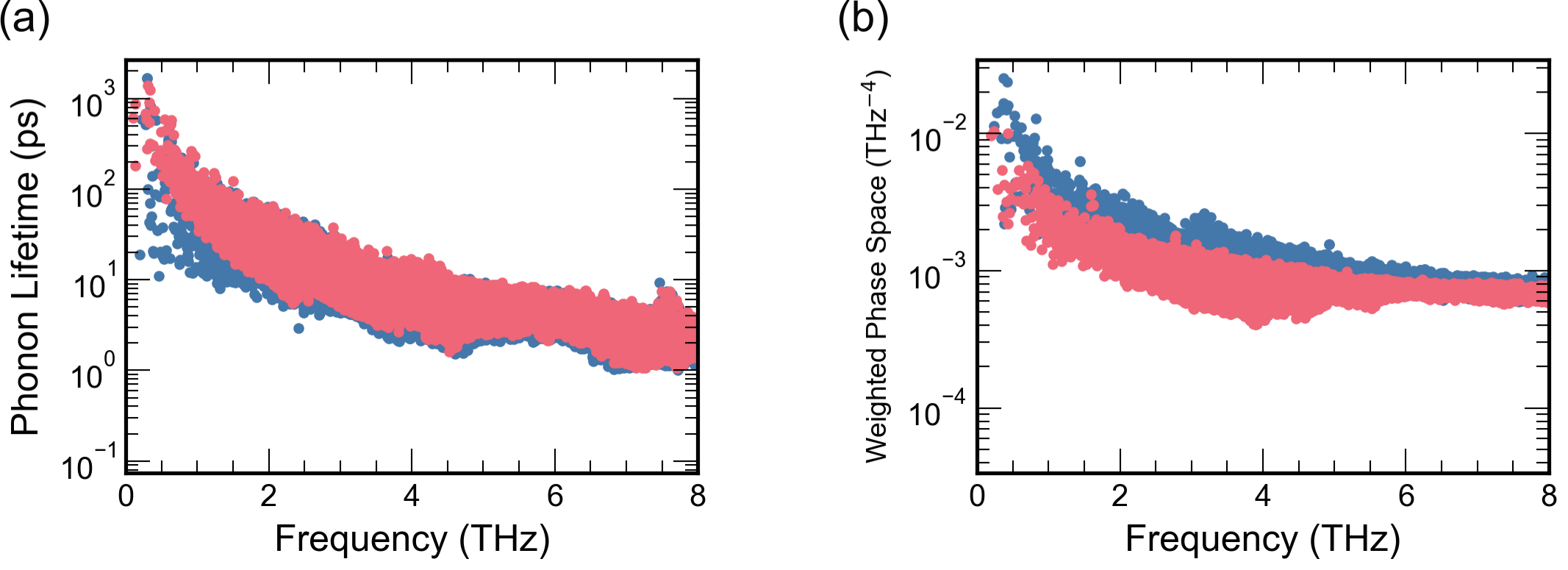}
\caption{Phonon lifetimes (a) and weighted phase-space (b) at 300 K for the bare IFC phonon model (blue) and the model including phonon renormalization at 300 K (red).}
\label{Fig_4}
\end{figure}

In \fig{Fig_4}(a), we plot the phonon lifetimes at 300 K for both the bare IFC (blue) and renormalized (red) phonons. We notice that the phonon lifetimes increase across the relevant phonon frequency range (0 - 8 THz). This is the origin of the observed \ce{\kappa} enhancement. Differences in the phonon lifetimes stem from either a change in the number of phonon scattering events that obey the energy and momentum conservation (\ie phase-space) or the higher-order IFCs. In \fig{Fig_4}(b), we compare the weighted phase-space from both models. The weighted phase-space represents the number of phonon interactions that obey energy and momentum conservation scaled by the occupation factors and normalized by the phonon frequencies. As a result, the weighted phase-space is much more sensitive to phonon frequency shifts compared to the phase-space alone (order $x^5$ vs. order $x$).\cite{li_2014a} It is clear that the weighted phase-space for the renormalized phonons is smaller than that of the bare phonons. This provides a clear answer to our first question. Accounting for phonon renormalization boosts the \ce{\kappa} as a result of longer phonon lifetimes due to a more restricted phonon phase-space.

To address the anisotropic \ce{\kappa} enhancement, we need to analyze the phonon group velocity as it is the only parameter in the \ce{\kappa} expression (Equation 1) that has directional components. For this purpose, we define a heat-averaged group velocity $\langle v^\alpha \rangle$ :
\begin{equation}
\label{Eq_6}
\langle v^\alpha \rangle = \frac{\sum\limits_{\lambda} \kappa^\alpha_{\lambda} \lvert v^\alpha_{\lambda} \lvert}{\sum\limits_{\lambda} \kappa^\alpha_{\lambda}}
\end{equation}
where $\alpha$ represents the Cartesian direction and $\lambda$ is the mode index. This metric allows us to meaningfully link changes in the group velocity to changes in the \ce{\kappa}. The difference between the thermal conductivity in the direction of the lattice vectors (\ce{\kappa_001}) and that in the Cartesian direction (\ce{\kappa_z_z}) is discussed in the Supplementary Material. In \fig{Fig_5}, we plot the relative change in \ce{\kappa^\alpha} against the relative change in $\langle v^\alpha \rangle$.  We observe that there is clear correspondence between the change in $\langle v^\alpha \rangle$ to the anisotropic \ce{\kappa} enhancement. The largest relative \ce{\kappa} increase is in the y-direction, which also has the largest relative change in $\langle v^\alpha \rangle$. Furthermore, the lowest \ce{\kappa} increase is in the z-direction, which has a slight decrease in $\langle v^\alpha \rangle$. The emergence of outliers is likely due to the non-trivial shifts in the phonon energies due to phonon renormalization.
The consistent observation of this trend across all studied temperatures underscores the generality of our analysis. This answers our second question\textemdash why is the increase in thermal conductivity anisotropic when accounting for phonon renormalization? With both questions addressed, the results are fully explained in the framework of the phonon gas model. Having demonstrated the importance of accounting for anharmonicity in the phonon energies, we now turn our attention to addressing the anharmonicity of the atoms in real space, with an attempt to link it to the \ce{\kappa} increase due to phonon renormalization. 
\begin{figure}[htbp]
\centering
\includegraphics[scale = 0.65]{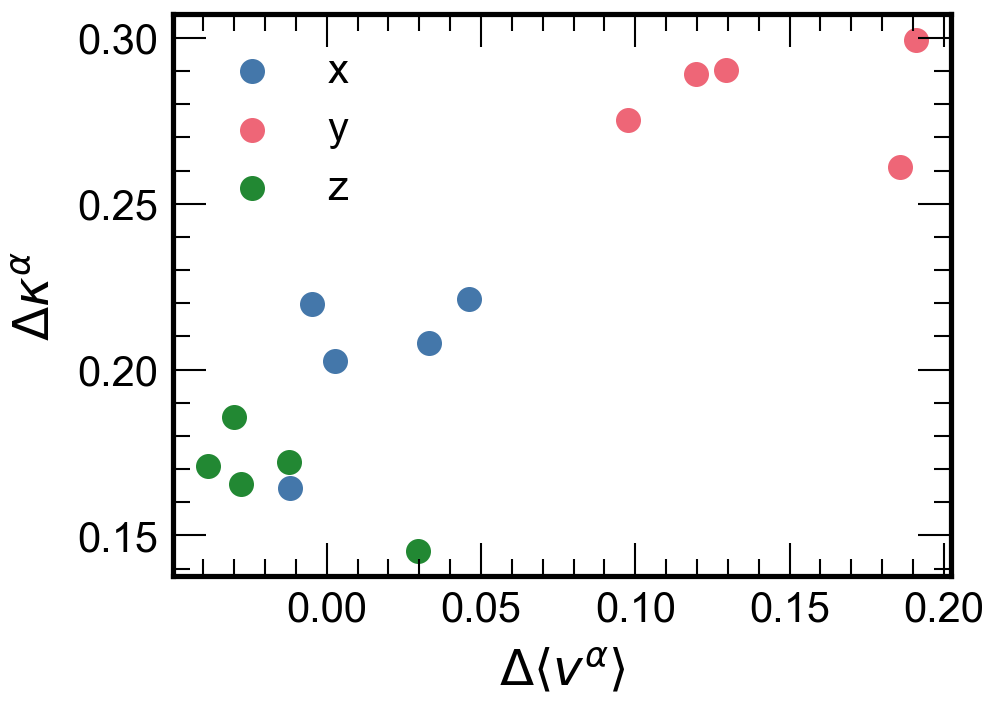}
\caption{Scatter plot of the relative change in thermal conductivity ($\Delta \kappa^\alpha$) against the relative change in the heat-averaged group velocity ($\Delta \langle v^\alpha \rangle$) for the three Cartesian directions: x (blue), y (red), and z (green). Note that the x and y-directions correspond to the (100) and (010) directions, respectively.  However, the z-direction does not correspond to a lattice vector owing to the non-orthogonal nature of the crystal. See Supplementary Material for more detail.}
\label{Fig_5}
\end{figure}

\begin{figure}[htbp]
\centering
\includegraphics[scale = 0.65]{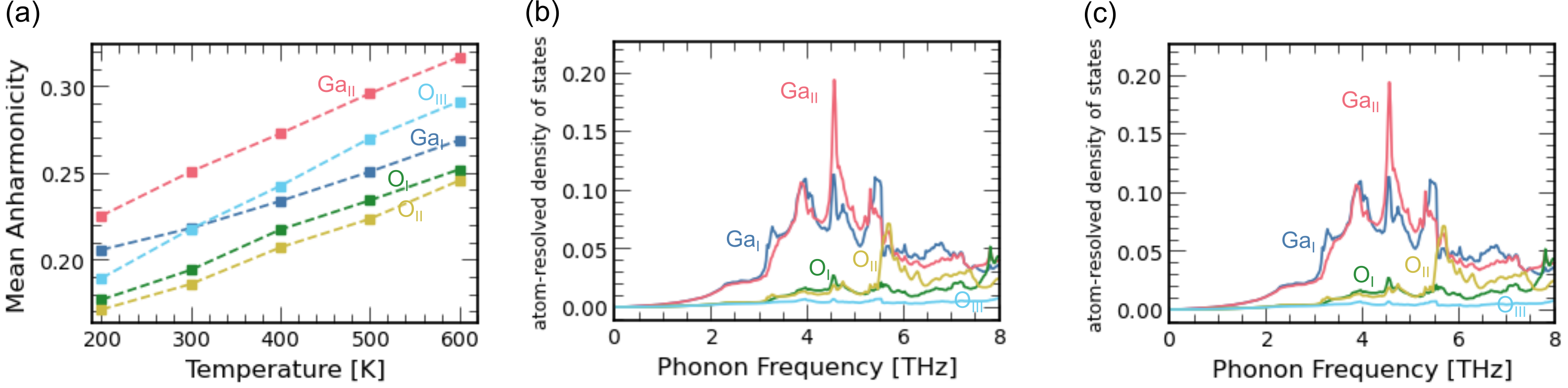}
\caption{Temperature-dependence of the non-equivalent atoms anharmonicity (a) and the atom-resolved phonon density of states at 300 K for the bare IFC (b) and renormalized (c) phonons.}
\label{Fig_6}
\end{figure}

Following the modified anharmonicity metric of Knoop \etal\cite{knoop_2020}, we plot the temperature-dependence of the mean anharmonicity for all unique atoms in \ce{\beta-Ga_2O_3} (\fig{Fig_6}(a)). Although the mean anharmonicity at 300 K of \ce{\beta-Ga_2O_3} (0.21) is only slightly higher than of silicon (0.15)\cite{knoop_2023}, examining the anharmonicity of the individual atoms reveals important physics. Most importantly, we observe that the octahedrally-coordinated \ce{Ga_I_I} (see \fig{Fig_1}) is the most anharmonic atom across the studied temperature range. We also plot atom-projected phonon density of states (PDOS) at 300 K for bare (\fig{Fig_6}(b)) and renormalized (\fig{Fig_6}(c)) phonons. It is clear that \ce{Ga_I_I} also peaks and has the highest PDOS contribution in the phonon frequency range where the \ce{\kappa} enhancement occurs (0 - 8 THz). The \ce{Ga_I_I} contribution to the PDOS even increases when we account for phonon renormalization. This strongly suggests that the \ce{\kappa} increase is directly related to the dynamics of the \ce{Ga_I_I} nuclei. This reveals the origin of the failure of the harmonic approximation to accurately predict the anisotropic \ce{\kappa} in \ce{\beta-Ga_2O_3}. Therefore, it can be argued that including phonon renormalization in the computational model leads to an increase in \ce{\kappa} because it has a more accurate description of the dynamics of the \ce{Ga_I_I} nuclei.

By examining the anharmonicity of the individual atoms in \ce{\beta-Ga_2O_3}, we notice that atoms of the same element have significantly different anharmonicities. In particular, we focus on tetrahedrally-coordinated \ce{Ga_I} and octahedrally-coordinated \ce{Ga_I_I}, as they have the highest contribution in the PDOS range of interest (0 - 8 THz). The fact that \ce{Ga_I} has lower anharmonicity than \ce{Ga_I_I} is in-line with results obtained by Xia \etal \cite{xia_2020} In their study, they calculate the \ce{\kappa} of a wide range of binary rocksalt (octahedral-coordination) and zinc-blende (tetrahedral-coordination) materials. They discover that regardless of a material's \ce{\kappa}, rocksalt materials have a stronger response (larger \ce{\kappa} increase) to phonon renormalization than zinc blende materials. However, due to \ce{\beta-Ga_2O_3}'s complex structure, gallium atoms with both coordinations exist. Therefore, what we reveal in our study about the link between atom anharmonicity, coordination, and phonon renormalization has been shown before, albeit for materials that have only a single coordination in their structure. In the final part of this study, we demonstrate how an analysis of atomic anharmonicity could be used to tailor vibrational properties in complex structures. 

The higher anharmonicity of \ce{Ga_I_I} relative to \ce{Ga_I} opens up an interesting question: what if we replace \ce{Ga_I_I} with another atom with lower anharmonicity with the purpose of increasing \ce{\kappa}? In Slack's seminal work\cite{slack_1973}, he proposes four rules for finding materials with higher thermal conductivity: low atomic mass, strong bonding, simple crystal structure, and low anharmonicity. Therefore, it would be reasonable to replace \ce{Ga_I_I} with boron, the lightest element in group III. However, it has been shown that it is thermodynamically favorable for boron to occupy the \ce{Ga_I} site\cite{lehtomaki_2020}. The second-lightest atom in the group is Aluminum. Substituting in aluminum, which preferentially occupies the \ce{Ga_I_I} site, to create a structured \ce{\beta-AlGaO_3} alloy has been shown to increase the mean \ce{\kappa} by more than 70\%.\cite{mu_2019} Here, we have demonstrated that using Slack's rule to lower the atomic masses, atom anharmonicity to decide on which element to replace, and thermodynamics to understand preferential occupation sites, provides a rational design process that enables tailoring the vibrational properties of complex materials. A comprehensive study to tune the thermal transport or optical response in \ce{\beta-Ga_2O_3}, or other complex structures, is the subject of a future study.

%--------------------------Conclusion-----------------------------------------------------------%
\section{Conclusions}

To conclude, using DFT coupled with a solution of the PBTE, we resolve the inconsistency between experimental results and computational models for \ce{\kappa} in \ce{\beta-Ga_2O_3}. Our findings reveal that phonon renormalization increases \ce{\kappa} anisotropically. This anisotropic enhancement is attributed to phonons with longer lifetimes due to a more restricted phase-space and anisotropic changes in phonon group velocities.
Moreover, our investigation into the anharmonicity of atoms identified the octahedrally-coordinated gallium atom as the most anharmonic, likely causing the harmonic phonon model's failure in 
 \ce{\beta-Ga_2O_3}. We also gave insights that atomic anharmonicities could serve as a valuable metric for tuning the vibrational properties of materials. These insights provide a pathway for optimizing thermal and optical properties in 
\ce{\beta-Ga_2O_3} and other complex materials.
%------------------------------------------------------------------------------------------------%

%--------------------------Data Availability Statement-------------------------------------------%
\section{Data Availability}
Code used to produce the figures in this manuscript and underlying data is available upon reasonable request to the corresponding author. 
%------------------------------------------------------------------------------------------------%

%--------------------------Acknowledgements--------------------------------------------------%
\section{Acknowledgements}
A.A. acknowledges financial support from Kuwait University. Z.H., Z.G., and X.R. acknowledge partial support from US National Science Foundation Award Nos. 2311848 and 2321301. A.A. thanks Olle Hellman for providing access to TDEP and Florian Knoop at Link\"{o}ping University for TDEP assistance and valuable discussions.

%------------------------------------------------------------------------------------------------%

%------------------------------------------------------------------------------------------------%
% The bibliography
\bibliography{Manuscript_V1}
%------------------------------------------------------------------------------------------------%

\end{document}